\documentclass[proof]{pasj00}
\pdfoutput=1
\usepackage{lscape}
\usepackage{color}
\draft

\begin{document}
\SetRunningHead{M. Kimura et al.}{Is the Cygnus Superbubble a Hypernova Remnant?} 
\Received{2012/07/31}
\Accepted{2012/09/13}

\title{Is the Cygnus Superbubble a Hypernova Remnant?}  

\author{  
	   Masashi \textsc{Kimura}\altaffilmark{1}, \\
           Hiroshi \textsc{Tsunemi}\altaffilmark{2},
	   Hiroshi \textsc{Tomida}\altaffilmark{1},
	   Mutsumi \textsc{Sugizaki}\altaffilmark{3}, \\
	   Shiro \textsc{Ueno}\altaffilmark{2},
	   Takanori  \textsc{Hanayama}\altaffilmark{4},
	   Koshiro  \textsc{Yoshidome}\altaffilmark{4},	   
	   Masayuki \textsc{Sasaki}\altaffilmark{2}
 }
 \altaffiltext{1}{ISS Science Project Office, ISAS, JAXA, 2-1-1 Sengen, Tsukuba, Ibaraki 305-8505, Japan
                  }
 \email{kimura.masashi@jaxa.jp}
  \altaffiltext{2}{Department of Earth and Space Science, Graduate School of 
                  Science, Osaka University, 1-1 Machikaneyama, Toyonaka, Osaka
                  560-0043, Japan
                  }
 \altaffiltext{3}{MAXI Team, Institute of Physical and Chemical Research (RIKEN), \\ 2-1 Hirosawa, Wako, Saitama 351-0198, Japan}

 \altaffiltext{4}{Department of Applied Physics, Faculty of Engineering, University of Miyazaki, \\ 1-1 Gakuen Kibana-dai
Nishi, Miyazaki, 889-2192, Japan}


%


\maketitle

\begin{abstract}
We present here the observation of the Cygnus Superbubble (CSB) using the Solid-state slit camera (SSC) aboard the
Monitor of All-sky X-ray Image. The CSB is a large diffuse structure in the Cygnus region with enhanced soft X-ray emission.
By utilizing the CCD spectral resolution of the SSC, we detect Fe, Ne, Mg emission lines from the CSB for the first time.
The best fit model implies thin hot plasma of  kT $\approx$ 0.3 keV with depleted abundance of $0.26 \pm 0.1$ solar.
Joint spectrum fitting of the ROSAT PSPC data and MAXI/SSC data enables us to measure precise values of $N_{\rm H}$ and temperature inside the CSB.
The results show that all of the regions in the CSB have similar $N_{\rm H}$ and temperature, indicating that the CSB is single unity.
The energy budgets calculation suggests that 2-3 Myrs of stellar wind from the Cyg OB2 is enough to power up the CSB, whereas due to its
off center position, the origin of the CSB is most likely a Hypernova.
\\
\end{abstract}

\section{Introduction}
In the vicinity of a massive star such as type O or B, either because of the strong wind or due to the explosion of the star, 
the surrounding interstellar medium (ISM) is blown off. The combine effects from stellar winds and supernova explosion from stellar association containing dozens of O and B stars can produce quite large cavities filled with hot gas that are called superbubbles.
Three giant bubbles have been identified so far in the local arm of the Milky Way. These structures are known as
the Orion-Eridanus (\cite{Reynolds1979}), the Scorpio-Centaurus (\cite{Weaver1979}) and the Cygnus superbubble (CSB, \cite{Cash1980}).
 
 The CSB is known for its strong X-ray emission and its large size, which is about $18^{\circ} \times 13^{\circ}$ along the Galactic longitude and latitude, respectively (\cite{uyaniker2001}). 
 \citet{uyaniker2001} also reported that no structures show up which resemble the horse shoe shaped X-ray structure in radio continuum.
 Although the CSB was discovered by the HEAO-1 X-ray observation (\cite{Cash1980}), very few studies have been done in X-ray due its large size.
 Numerous OB associations at different distances are located in its direction, whereas their physical connections to the CSB are still unclear. 
 
Previous studies suggest various theories about the formation of this superbubble,
 such as sequential explosions of several tens of supernovae in Cyg OB2 (\cite{Cash1980}).
 They estimated present internal energy $E_{T}>6\times 10^{51}$ergs and assumed that the CSB is related to Cyg OB2 therefore its distance is $\sim 2$kpc.
 To create such structure with energy and size (450 pc in diameter), its initial energy have to be E$_0 \simeq 10^{54}$ergs.
 Since this energy is no less than three orders of magnitude greater than the energy output of a single conventional supernova, 
 they concluded with a hypothesis that the CSB was produced by a chain of 30-100 conventional supernova explosions over the past 3-10 million years.
 
 \citet{abbott1981}, on the other hand, suggested that the CSB was produced from strong stellar winds flowing from stars in the Cygnus OB2 association.
 Cygnus OB2 is a compact star group that creates one of the strongest stellar winds in the Galaxy: it contains at least 3000 stars and about 300 of them are OB stars (\cite{reddish1966},\, \cite{humphreys1978}).
Another interpretation done by  \citet{blinnikov1982} is that the CSB is the remnant of an explosion of a single super massive star with the energy of $10^{52-53}$ ergs.
\citet{iwamoto1998} observed a supernova with explosion energy of $ \sim 2 - 5 \times 10^{52}$ ergs and called this phenomenon a hypernova.
They stated that the hypernova is an explosion of massive progenitor star of $\sim 40{\rm M}\solar$. Since this explosion energy is very similar to the prediction by \citet{blinnikov1982}, the CSB might be a hypernova remanent.

 Another explanation done by  \citet{uyaniker2001} is that the CSB is not a physical unity. 
 It consists of a superposition of multiple components such as supernova remnants, shells around individual stars and OB associations swept up by the wind. They are scattered in the local spiral arm at the distance from 1 to 5 kpc.  Therefore, they claim that the CSB is not a physical unity.

 

\section{Observation and Data reduction}


\subsection{MAXI/SSC Calibration and Data reduction}
 The Monitor of All-sky X-ray Image (MAXI, \cite{matsuoka2009}) is the astronomical payload aboard the international space station (ISS).
The MAXI has two types of cameras: the Gas Slit Camera (GSC, \cite{mihara2010}, \cite{sugizaki2011}) and the Solid-state Slit Camera (SSC, \cite{tsunemi2010}, \cite{tomida2011}).
The MAXI/SSC consists of two identical cameras.  Since it has no X-ray collecting devices, it observes the sky with two FOV of $1^\circ.5 \times 90^\circ$ aimed at the $16^\circ$ above the earth-horizon and the zenith directions.  The SSC allows us to scan all sky in the energy range of $0.5-12.0$ keV with the spectral resolution of the X-ray CCD, this makes the  SSC a suitable instrument to study a large diffuse X-ray structure such as the CSB.

The SSC in-orbit performance and the basic calibration are explained in \cite{tsunemi2010}. 
The PHA-energy gain calibration of the CCD's is done by using Cu emission line in the background originating from the collimator. 
The temperatures of CCDs vary by about 10$^\circ$C during each ISS orbit of about 92 minutes.
Since the CCD gain showed the variation correlated with the temperature, we used this correlation to correct the CCD-gain variation.

We used the data taken from 2009 August 18 to 2012 February 01.  
In order to acquire clean event data, we applied several event selection methods.
The frame image taken by the MAXI/SSC when the ISS is on the day-time earth suffers from IR/visible light contamination.  It causes saturation near the edge of each CCD. 
Since these data are not applicable to the spectrum analysis, we excluded them in the following analysis.

The moon also becomes another source of IR/visible light. 
The data taken from radius of R$<10^{\circ}$ from the moon are
affected low energy background of the SSC, hence we mask these regions, and exclude them from our data analysis.
Although the SSC does not observe the sky when it passes through the South Atlantic Anomaly (SAA), it does observe the sky when it is in high latitude. The data taken when it is in latitude higher than $40^\circ$ show very high background,  therefore we excluded these data from our analysis.
After applying these selections, our exposure time became about $1.1 \times 10^6$ seconds.

We also applied grade selection method. Figure 6 in \citet{tsunemi2010} shows the SSC background spectrum in various grade selections.
The G0 represents the single pixel event and G1 and G2 represents the split pixel events.
The grade allocation is done on the on-board software.
The G0 spectra are background contaminated below 0.7 keV that depends on the working temperature of the SSC.  
We select the data above 0.7 keV so that we can have good data.
The G1+G2 spectrum in this figure shows a large peak around 0.7 keV that must come from the combination between the event threshold and the split threshold.  
Since this peak does not appear on G0 spectrum,  we employ only G0 event in the energy range of 0.7-1.7 keV and employ G0, G1, G2 event for 1.7-7.0 keV. 
We divide energy range below and above 1.7 keV (Si-K edge) because the quantum efficiency of the CCD changes at 1.7 keV.
We exclude the data above 7.0 keV since the non X-ray background (NXB) become dominant. 
Therefore, the effective energy range in our analysis is 0.7-7.0 keV.

\subsection{All sky image}
\begin{figure}
  \begin{center}
\includegraphics*[width=160mm, bb=0 0 1516 771]{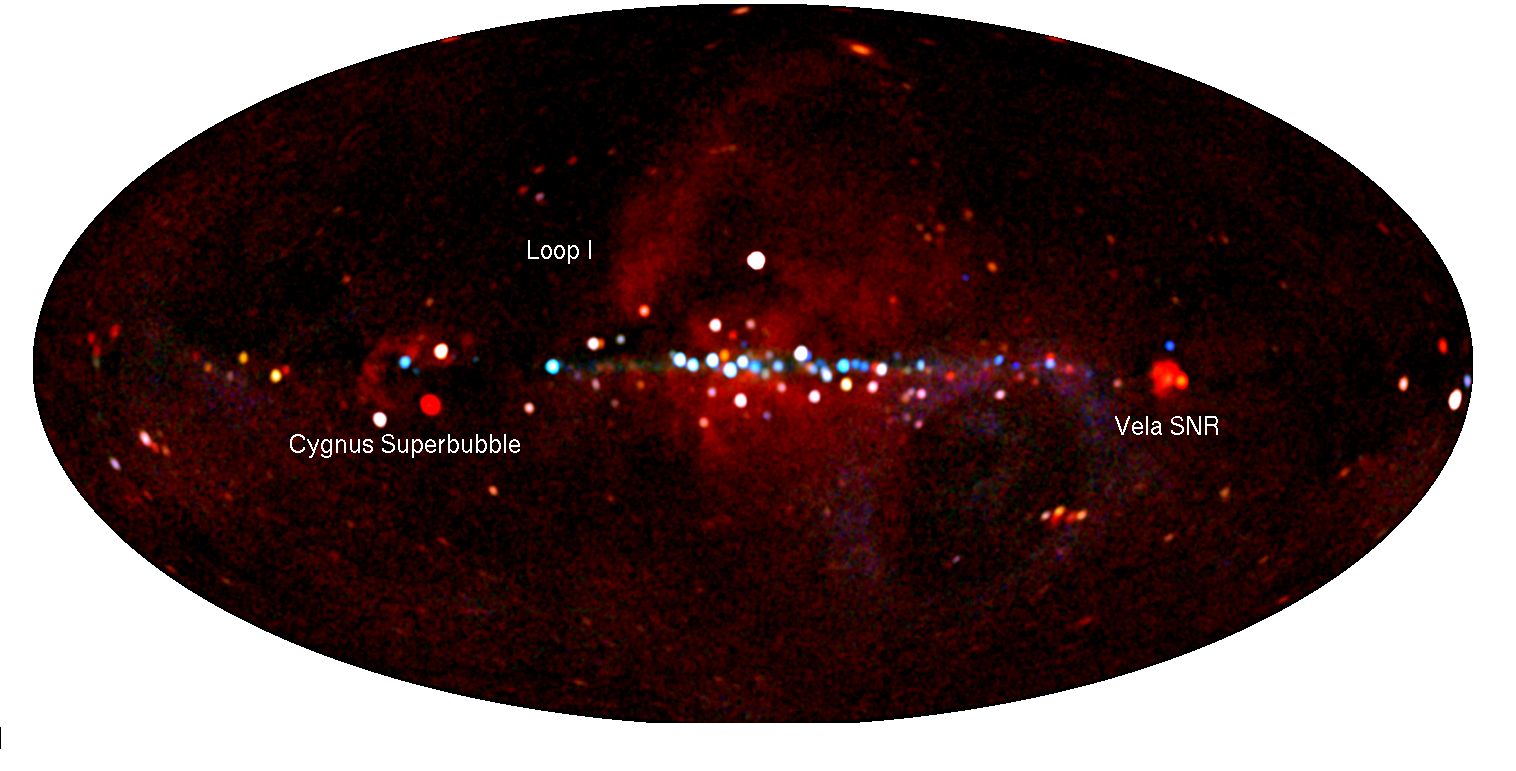}
\end{center}
  \caption{All sky image obtained by the SSC for 30 months. Background is not subtracted}\label{AllSkyImage}
\end{figure}

 Figure \ref{AllSkyImage} shows the all sky map created from the SSC data after event selection described in the previous subsection.
The red, green and blue on the map correspond to the energy bands of 0.7-1.7 keV, 1.7-4.0 keV and 4.0-7.0\,keV, respectively.  This map is exposure-time corrected, while no background is subtracted.  The map shows over 140 point sources, and several large-scale structures can also be seen such as the CSB, the Loop-I and the Vela SNR.

\section{Background}
The background study is important, particularly in analyzing diffuse structures. 
The conventional method such as taking annular region around the CSB did not give us good statistics.
Since we want good statistics in background spectrum, we employes all data taken by MAXI/SSC to estimate the averaged background spectrum.

The background for extended sources mainly consists of four components (\cite{miller2008}).  They are the cosmic X-ray background (CXB), the Galactic Halo Component (GH), the Local hot bubble (LHB) and the Non-X-ray background (NXB). Among them, the LHB does not play an important role in the SSC energy band.  The CXB and the GH are temporally stable components while their spatial uniformity is well studied (\cite{tawa2008}).  With taking into account the point spread function (PSF) of the SSC, their uniformity is about 3\% in intensity.  In most of the X-ray observatories, the NXB can be measured by employing the data looking at the night earth.  In the case of Suzaku/XIS, the NXB is sorted according to the cut-off-rigidity (COR) with good statistics.  On the contrary, the SSC never sees the night earth with the exception of the period when the ISS moves upside down due to the special maneuvers such as the space craft docking.  So far, we have not obtained enough amount of the NXB data.  Therefore, we take a special method for the estimation of the SSC background.  First of all, we selected the data explained in the previous section.  In this section, we will select the sky regions where there is no bright X-ray sources.  Then we will sort them by the COR.  In this way, we will estimate the SSC background (NXB+CXB+GH) for diffuse sources.

\subsection{All sky map by using all good data}

\begin{figure}
  \begin{center}
\includegraphics*[width=16cm, bb=0 0 1463 770]{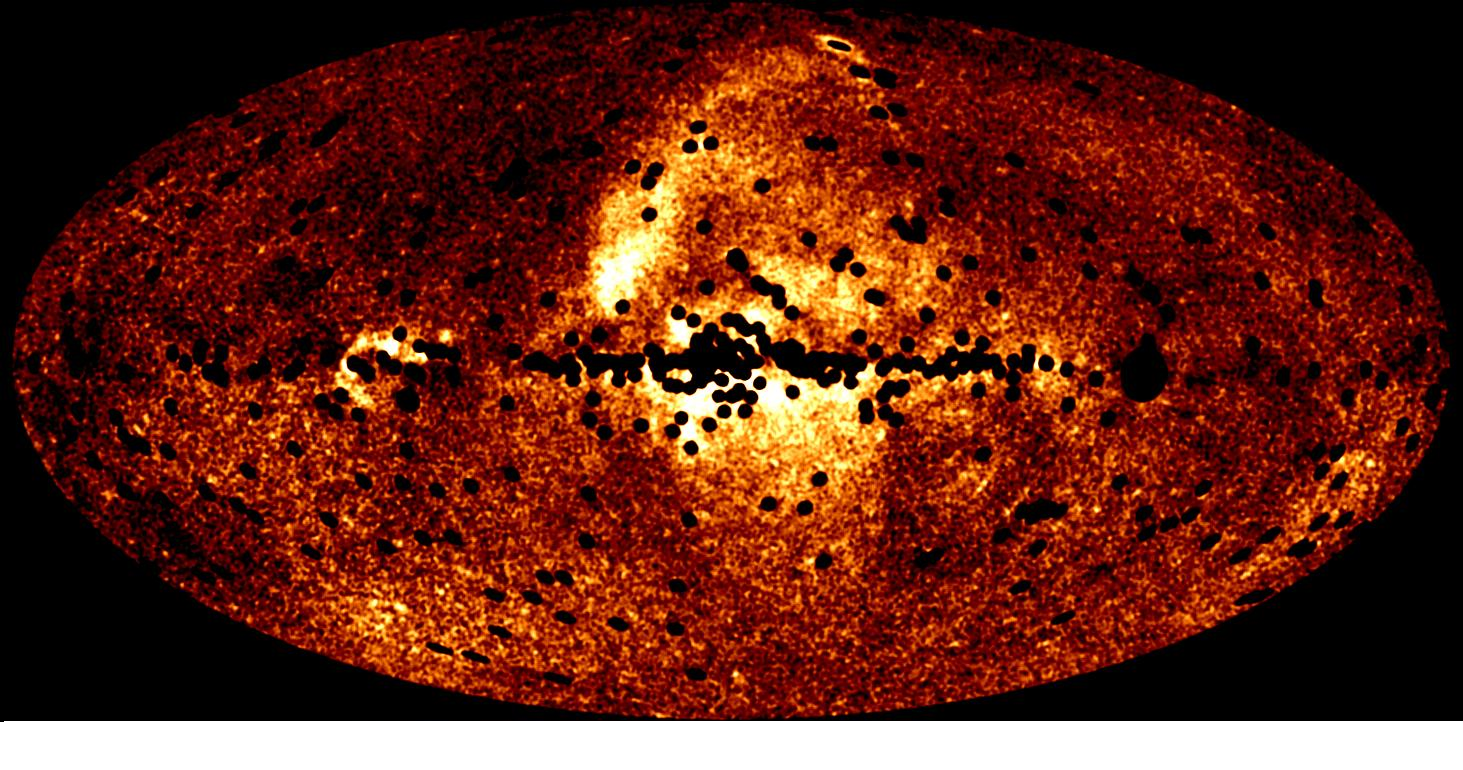}
\end{center}
  \caption{Low band (0.7-1.7keV) all sky image without point sources. Background is not subtracted.}\label{no_source}
\end{figure}

We mainly focused on estimating the background of low energy band (0.7-1.7 keV) since most of the diffuse structures can be seen in this energy range. 
First, we took out all the known point sources from the SSC data.  Figure \ref{no_source} shows all sky image in the low energy band without point sources. The list of point sources are taken from MAXI/GSC public data (http://maxi.riken.jp), which includes point sources brighter than 3 mCrab in 2 -- 10 keV band.

The radius of $1^{\circ}.5$ is used to take out point sources, but the radius of $3^{\circ}.0$ and $5^{\circ}.0$ are applied for two bright SNRs, the Cygnus loop and the Vela SNR.  

\begin{figure}
  \begin{center}
\includegraphics*[width=10cm, bb=0 0 596 372]{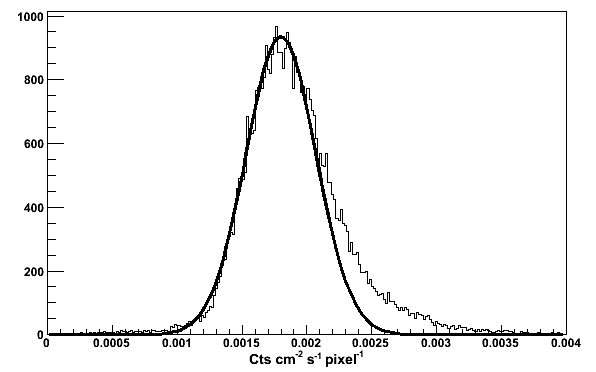}
\end{center}
  \caption{Flux histogram of low energy band (0.7-1.7keV). }\label{low_hist}
\end{figure}

Now that the point sources are taken out, we split all sky into 49152 pixels using Healpix (\cite{gorski2005}).  Next, we calculated the count rate of each pixel.
Figure \ref{low_hist} shows the count rate histogram of the 0.7-1.7 keV energy range, showing an asymmetric distribution.  The peak corresponds to the average of the SSC background.  The lower side shows the background structure while the higher side shows the galactic diffuse components.  
We find that the lower side of the data can be expressed by a gaussian function.  
Although each pixel have different exposure time, it contains about 300 photons in average.
Therefore statistical error is about 6\%.
However, $\sigma$ of the best fit Gaussian is about 15\%, indicating that it contains some systematic uncertainties. 
Since we are estimating the background for diffuse sources, we removed the pixel with count rate higher or lower than 2$\sigma$ of the gaussian function. Finally, we considered emission from remaining pixel to be background which should contain both  NXB and CXB.

\subsection{SSC background based on the COR}
\begin{figure}
  \begin{center}
\includegraphics*[width=10cm, bb=0 0 760 556]{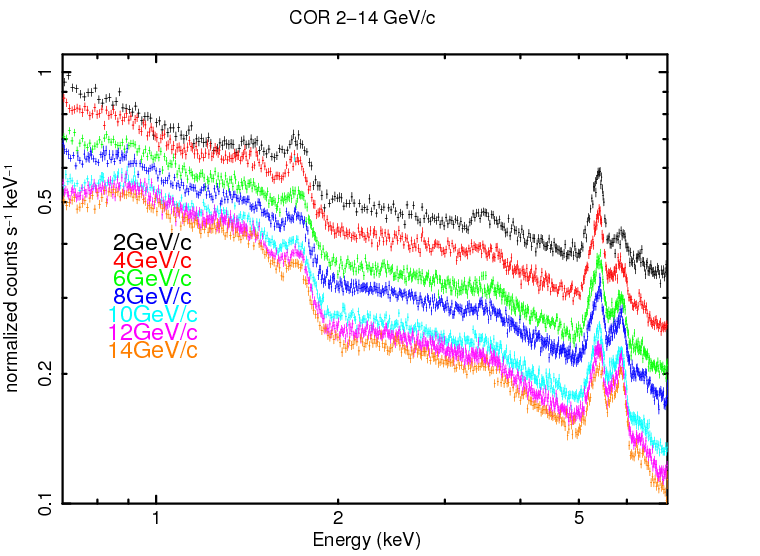}
\end{center}
  \caption{Background spectra for each CORs.  This figure shows 7 spectra out of 13 for simplicity. }\label{cor1}
\end{figure}

The background of the SSC mainly comes from the NXB which is caused by charged particles and $\gamma$-rays entering the detector from various directions. Therefore, the NXB varies with time according to the radiation environment of the ISS, which must be a cause of the systematic uncertainties. 
This is strongly correlated with the COR.  The NXB+CXB event file created earlier have 
detection time when the MAXI/SSC was in various COR, so we split the event file according to the COR.
The MAXI/SSC is operated when its COR is in the range of 2--14GeV/c, we split the event file into 13 pieces, this will be our background database. 
Figure \ref{cor1} shows the spectrum of NXB+CXB for each CORs. The emission line around 1.7 keV comes from the Si in CCD and the emission line in 5.5 keV comes from the Cr used in SSC's body.  As the flux of the background obviously varies according to the COR, there is about twice of differences between COR of 2 GeV/c and 14 GeV/c.
 
In this way,  we know the flux of background in different COR.  In order to make background spectra for the target object,  
we find the ratio of exposure time in 13 different CORs.
Then, we can calculate the weighted average of the background database according to the COR.  
This will be the background spectrum for target object.
In order to check the correctness of our background, we compared spectra of Cassiopeia A obtained by SSC and Suzaku/XIS \citep{koyama2007} and confirmed that they are consistent with each other. Both Cas A and CSB are located near Galactic plane but far away from galactic center
($l_{CSB} = 70^\circ \sim 95^\circ$,where $l_{Cas A}=110^\circ$) therefore emission from Galactic ridge X-ray emission should be negligible.

\begin{table}[hbt]
\begin{center}
\caption{\sc The best-fit gaussian parameters of figure \ref{low_hist}} 
\label{bg_fit}
\begin{tabular}{ccc} \hline\hline
Component&Parameters & Value \\
\hline 
Gaussian & Center      (cts sec$^{-1}$cm$^{-2}$pixel$^{-1}$) & $(1.827 \pm 0.002) \times 10^{-3}$ \\
&$\sigma$  (cts sec$^{-1}$cm$^{-2}$pixel$^{-1}$) & $ (2.82 \pm 0.02)\times 10^{-4}$ \\
\hline
\end{tabular}
\end{center}
\end{table}

\section {Data analysis}
\subsection{ MAXI/SSC Data Analysis}
Figure \ref{csbImage} shows the zoomed image of the Cygnus region obtained by the SSC.
Some point sources along with the \lq\lq horse-shoe" shaped CSB are detected. 
For our analysis we excluded the three bright point sources in our field of view, which are Cygnus X-1, X-2 and X-3.
After masking these point sources we extracted spectrum from the entire horse-shoe region of the CSB.
Figure \ref{csb_spec} is the spectrum extracted from the region showed in the white lines in figure \ref{csbImage}.
The red mark on figure \ref{csb_spec} (left) shows the background spectrum based on the previous section where the black mark shows the source spectrum including background. 

Figure \ref{csb_spec} (right) is the background subtracted spectrum of the CSB. 
Since it showed a clear emission line around 1.3 keV, we fitted the spectrum with an absorbed bremsstrahlung model with several gaussian functions for emission lines.
We fix the width parameter of three gaussian functions to be 0.
We used XSPEC v12.6.0 for spectral fitting. 
It turns out that three gaussian functions are required to fit the spectrum properly. 
By adding three gaussian models, the fit improve to $\chi^2/{\rm dof}=189/148$, where simple absorbed bremsstrahlung model gives $\chi^2/{\rm dof}=251/154$
while the dof stands for the degree of freedom. 

Table \ref{ssc_fit} shows line center energies and the equivalent widths (EW) of the best fit.
Judging from the center energies of the three line emissions, the gaussian 1, 2 and 3 are emission lines from Fe-L, Ne-IX, and Mg-XI respectively, indicating that the 
spectrum has indeed thin-thermal origin.
We also tried to fit the spectrum with thin-thermal plasma models such as NEI and CIE.
The model which gave us the best fit was {\bf phabs*apec} (\cite{smith2001}) with $\chi^2/{\rm dof}=198/154$. 
We used the solar abundances by \citet{anders1989} and photoelectric absorption cross sections by \citet{balucinska1992}.
The parameters are also in table  \ref{ssc_fit}.  We were able to obtain the precise value of abundance of the CSB for the first time. 

In order to get the best fit, we added extra power-law model with $\Gamma = 1.41$ (\cite{kushiro2002}) to represent CXB component.
Although our background model includes the CXB component, \citet{tawa2008} showed that the flux of the CXB in Suzaku/XIS fluctuates about 14\% within its FOV that is 0.088 deg$^2$.
Since the PSF of the MAXI/SSC is about $1^{\circ}.5$, the CXB should fluctuates about 3\%. 
The best fit flux of additional power-law component is $0.5 \pm 0.3\%$ of CXB, which can be explained by the fluctuation of the CXB. 
We can conclude that the source spectrum and background spectrum
agrees well with each other in the energy band above 2 keV, meaning that we could not detect emission from the CSB above 2 keV.

\subsection{ ROSAT Data Analysis}
\citet{uyaniker2001} suggested that the CSB is  a group of diffuse sources located in different distances, 
so we split CSB into  five regions named by \citet{uyaniker2001}, {\bf Cyg-X}, {\bf North-East}, {\bf East}, {\bf S-ARC1}, {\bf S-ARC2}. 
They suggested that each region is located in different distance and that CSB is not physical coherent structure.  
However, they did not measure the values of absorption hydrogen column density ($N_{\rm H}$) by using X-ray data.
In order to estimate the distance to each region, we need to carefully measure the $N_{\rm H}$.
Although the SSC spectrum yields good plasma abundance, we cannot obtain precise value on its  $N_{\rm H}$. 
ROSAT (\cite{snowden1997}) has a high sensitivity in the soft X-ray band below 2 keV.
Therefore, we check the ROSAT PSPC-C scanning mode data to obtain precise values of $N_{\rm H}$.

\begin{figure}
\begin{tabular}{cc}
\begin{minipage}{0.51\hsize}
  \begin{center}
\includegraphics*[width=9cm, bb=0 0 717 564]{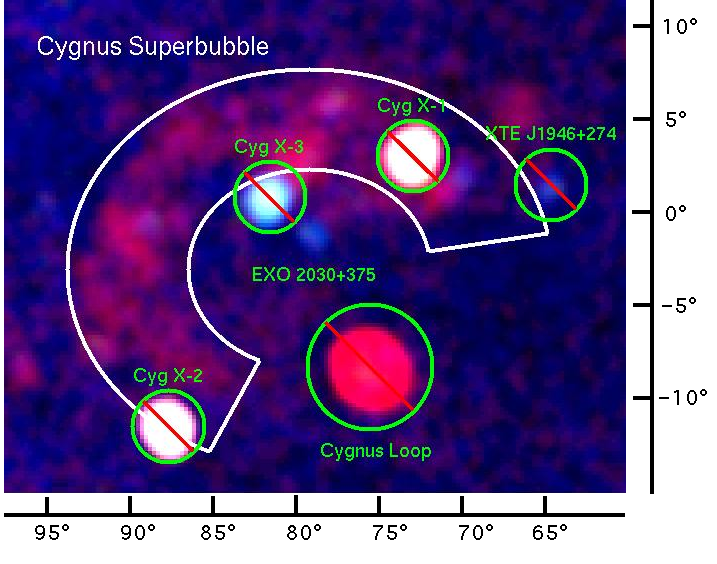}
  \end{center}
\end{minipage}

\begin{minipage}{0.51\hsize}
  \begin{center}
\includegraphics*[width=9cm, bb=0 0 717 564]{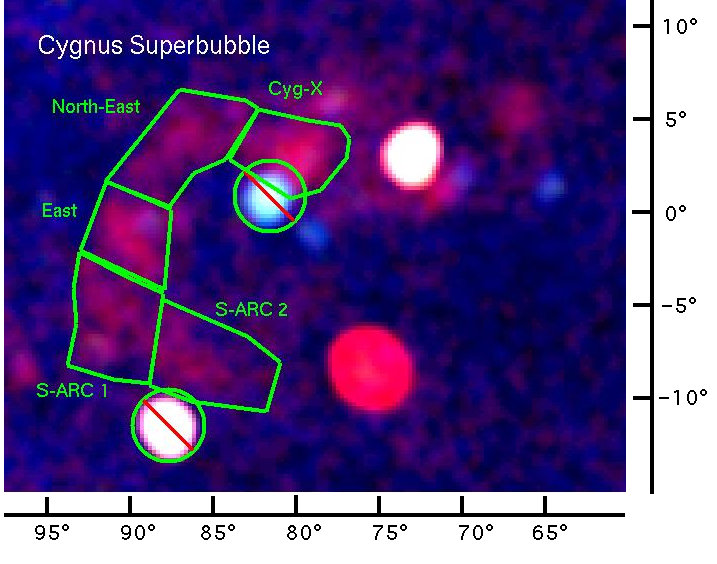}
    \end{center}
 \end{minipage}
\end{tabular}
      \caption{Left:zoomed image of the CSB. The horse shoe shaped region in white is the CSB. Several point sources are masked to exclude the emission. Right: same as left figure but with regions named by \citet{uyaniker2001}}\label{csbImage}
\end{figure}

\begin{figure}
\begin{tabular}{cc}
\begin{minipage}{0.5\hsize}
  \begin{center}
\includegraphics*[width=9.5cm, bb=0 0 761 504]{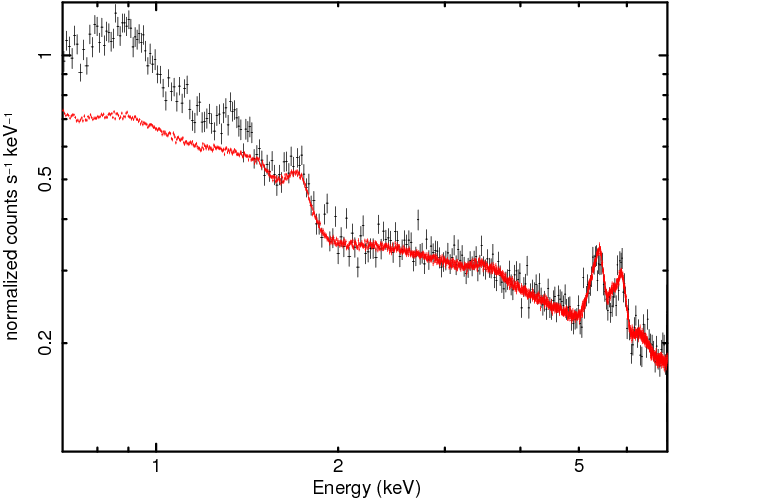}
  \end{center}
\end{minipage}

\begin{minipage}{0.5\hsize}
  \begin{center}
\includegraphics*[width=9.0cm, bb=0 0 713 498]{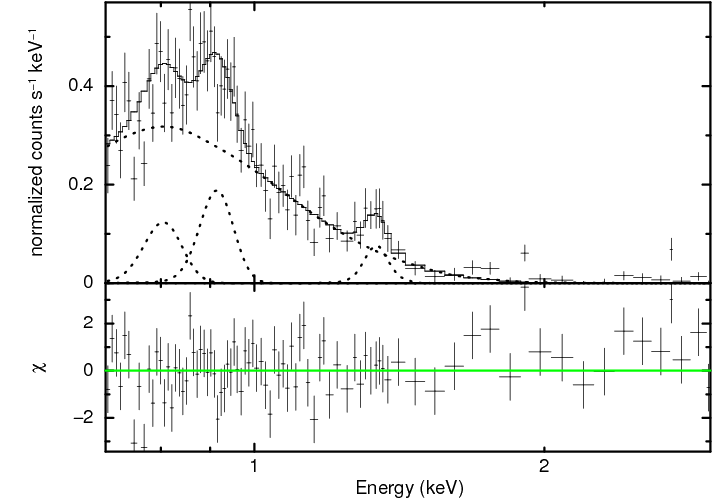}

    \end{center}
 \end{minipage}
\end{tabular}
      \caption{Left: Spectrum from entire CSB (black) and its background spectrum (red). 
      Right:Background  subtracted spectra of entire the CSB, fitted with an absorbed bremsstrahlung model with three gaussian functions. }\label{csb_spec}
\end{figure}

\begin{figure}
  \begin{center}
\includegraphics*[width=100mm, bb=0 0 533 384]{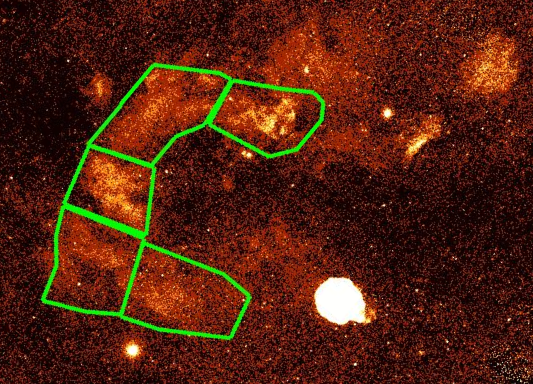}
\end{center}
  \caption{Image of the CSB from the ROSAT data (0.1-2.0keV)}\label{rosatimage}
\end{figure}

Figure \ref{rosatimage} is the image of the CSB from the ROSAT data in the energy range of 0.1-2.0 keV. 
The green regions in figure \ref{rosatimage} are the same region shown in figure \ref{csbImage} (right).
In order to perform consistent analysis with that by the MAXI/SSC, we only exclude three point sources just as we did in the SSC spectrum analysis.

\begin{figure}
\begin{tabular}{cc}
\begin{minipage}{0.5\hsize}
  \begin{center}
\includegraphics*[width=8cm, bb = 0 0 714 498]{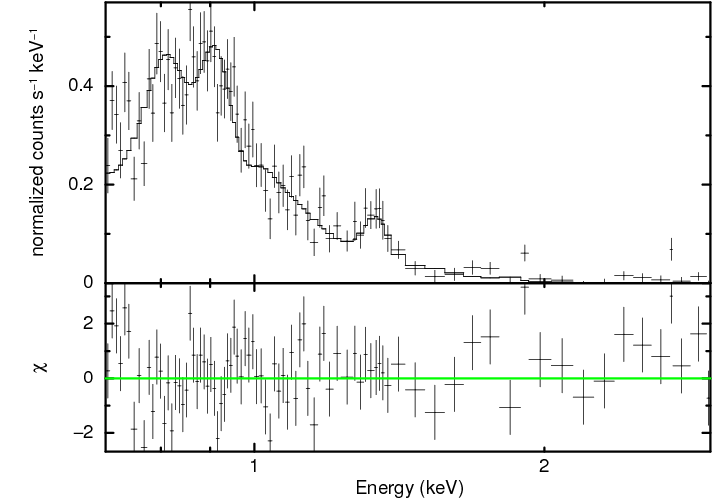}
  \end{center}
\end{minipage}

\begin{minipage}{0.5\hsize}
  \begin{center}
\includegraphics*[width=8.5cm, bb= 0 0 1650 1275]{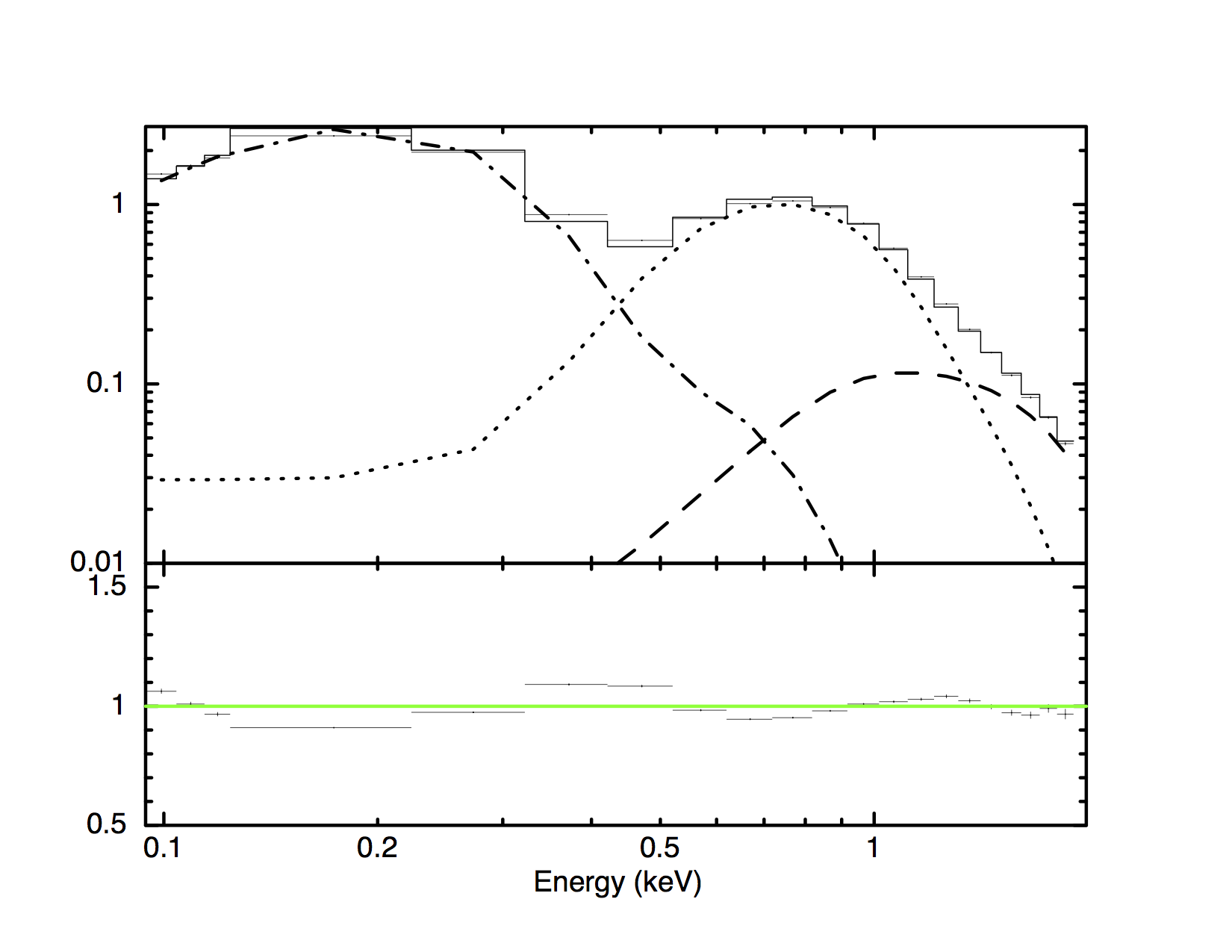}
    \end{center}
 \end{minipage}
\end{tabular}
      \caption{Left: Background  subtracted spectra of the entire CSB, fitted with {\bf phabs*apec} model.  Right:Spectra of the CSB obtained from the ROSAT. The dash-dotted line shows the LHB component, the dash line shows the CXB component and the dot line shows the CSB component which has identical parameters to spectrum in left panel. }\label{with_rosat}
\end{figure}

Figure \ref{with_rosat} (right) shows the spectra of the CSB obtained by the ROSAT. 
We subtracted the NXB from the ROSAT spectrum using the method explained by \citet{pluncinsky1993}.
Since we only subtracted the NXB component for the ROSAT spectrum, we added 2 background models: a CXB component and a LHB component.
The CXB component is shown in the dash line in figure \ref{with_rosat} , and it has the model of {\bf phabs*power-law} with $\Gamma = 1.41$ (\cite{kushiro2002}). 
The LHB component is shown in the dash-dot line in figure \ref{with_rosat} , 
and we used unabsorbed {\bf apec} model with kT = 0.1 keV and solar abundance (\cite{snowden1997}).
First, we re-fitted the spectrum from the entire CSB using the SSC and the ROSAT data with  {\bf phabs*apec}.
The parameters of this fit agree well with parameters of those of the SSC fit alone within the statistical uncertainty of the 90\% confidence limit.
Next we fitted the spectrum of five different regions obtained by the SSC and the ROSAT with the same model, 
we fixed the abundance value at that we obtained from the entire CSB spectrum, which is 0.26.
The spectrum fitting was performed by adding 2$\%$ systematic error to the ROSAT data due to its too good statistics (ROSAT User's Handbook, page 44).
Table \ref{fixednh_parameter} shows the values of $N_{\rm H}$ and temperature for each region.


\begin{table}[hbt]
\begin{center}
\caption{\sc SSC Spectrum fit of CSB} 
\label{ssc_fit}
\begin{tabular}{cccc} \hline\hline
Component & Parameters & Value \\
\hline 
Absorption &  $N_{\rm H}$ $(10^{22}{\rm cm}^{-2})$&  $0.30\pm0.08$\\
Bremsstrahlung &  kT(keV) &  $ 0.24 ^{+0.03}_{-0.08}$\\
Gaussian 1& $E_{center}$(keV)     &0.81$\pm$ 0.04 \\
&EW (keV)  &$0.04 \pm 0.03$ \\

Gaussian 2& $E_{center}$ (keV)     &0.93$\pm$ 0.03 \\
&EW (keV) &$0.06 \pm 0.03 $\\

Gaussian 3& $E_{center}$(keV)     &1.34$\pm$ 0.01 \\
&EW(keV)  & $0.11 \pm 0.01$\\
\hline
Absorption &  $N_{\rm H}$ $(10^{22}{\rm cm}^{-2})$& $0.32 \pm0.05$ \\
APEC          &  kT(keV)       &  $0.22 ^{+0.03}_{-0.01}$ \\
                  &  Abundance &  $0.26  \pm 0.1 $\\
\hline
\end{tabular}
\end{center}
\end{table}

\begin{table}[hbt]
\begin{center}
\caption{\sc ROSAT and SSC Spectra fit parameter of CSB} 
\label{fixednh_parameter}
\begin{tabular}{cccc} \hline\hline
                        & $N_{\rm H}$                       & kT     & Abundance           \\
 Region name  &$(10^{22}{\rm cm}^{-2})$     & (keV) & Relative to solar \\
\hline
All                   &   $0.30^{+0.01}_{-0.01}$   &    $0.228^{+0.007}_{-0.007}$   & $0.26^{+0.1}_{-0.1}$   \\
Cyg-X            & $0.27^{+0.04}_{-0.04}$&  $0.21^{+0.03}_{-0.03}$ & 0.26 (fixed)\\
East              & $0.28^{+0.03}_{-0.03} $&  $0.23^{+0.03}_{-0.03}$ & 0.26 (fixed)  \\
North-East     & $0.33^{+0.03}_{-0.03}$ &  $0.20^{+0.02}_{-0.02}$& 0.26 (fixed) \\
S-ARC1         & $0.23^{+0.03}_{-0.03} $&$0.22^{+0.02}_{-0.02}$ & 0.26 (fixed) \\
S-ARC2         &$0.22^{+0.03}_{-0.03}  $& $0.22^{+0.02}_{-0.02}$& 0.26 (fixed) \\
\hline
\end{tabular}
\end{center}
\end{table}



 \section{Discussion} 
The SSC spectrum of the CSB shows obvious Mg emission line and the spectrum can be reproduced 
by {\bf apec} model with relative abundance of 0.26. 
Although this value is significantly low compared to the solar abundance, similar abundance is observed in nearby SNR such as the Cygnus loop.
The distance to the Cygnus Loop is closer (540 pc , \cite{blair2005}) than that to the CSB, while it is in the same local arm as the CSB,  and its typical ISM abundance is about $\sim$ 0.3 (\cite{uchida2009}). 
Therefore, the emission from the CSB is most likely the swept-up ISM of the Cygnus region.

By model fitting of the MAXI/SSC and the ROSAT/PSPC spectra, we are able to get precise values of $N_{\rm H}$ and temperature
 of 5 different regions in the the CSB.
It turns out that in table \ref{fixednh_parameter}, neither temperature nor $N_{\rm H}$ show large difference in each region.
The value of $N_{\rm H}$ show small difference in each region, for example, region {\bf S-ARC1} and {\bf S-ARC2} showed
smaller $N_{\rm H}$ compare to other 3 regions. This is probably because each region has different angular distance from the Galactic plane.
{\bf S-ARC1} and {\bf S-ARC2} are far from the Galactic plane compared to other 3 regions, resulting  smaller values of  $N_{\rm H}$.
Although \citet{uyaniker2001} predicted that these 5 regions are in the same direction while they scatter in the line of sight from 1 to 5 kpc,
our observation indicates otherwise.
In this way, we conclude that the CSB is a physical unity.
\citet{yoshida2011} observed several stars in Cygnus OB2 using Suzaku satellite. 
Obtained spectra were well fitted by two-temperature CIE model with $N_{\rm H} = (0.2-0.4) \times 10^{22} {\rm cm}^{-2}$.
This clearly indicates that the CSB is well correlated with the Cygnus OB2, suggesting the CSB is in the vicinity of Cygnus OB2.

Since we were able to confirm that the CSB is thin-thermal and obtained its temperature and abundance, 
we can determine several parameters of the CSB to reveal its origin.
From the emission measure (EM: $\int n_en_Hd{\textrm V}$) of the model we calculated the density and the pressure of the CSB.
 In order to obtain these parameters, we have to estimate the size and the depth of the CSB. 
 We assume that the CSB is a single unity located near Cyg OB2 association, which is about 1.7 kpc away. 
 We guessed that the CSB has shape of $2/3$ of torus where R$_{\rm in} =5^{\circ}$ and R$_{\rm out} =11^{\circ}$, which leads its radius to be 330 pc
 (2 dimensional image of this torus is drawn in white line  in figure \ref{csbImage}). Assuming these dimensions, we calculated the volume of the CSB  as $4 \times 10^{62}$ ${\rm cm}^{3}$.  With EM and the volume of the CSB, we can calculate the density n$_e\sqrt{f}$ = 0.02 ${\rm cm}^{-3}$ , where $f$ is the filling factor.
Now that we know the density and temperature, the pressure of the gas is given below.
 \begin{equation}
  P/k = (n_e + n_{\rm ion})f^{-1/2}T = 1.1 \times 10^{5} {\rm cm}^{-3}{\rm K}
\end{equation}
While the pressure of the Galactic ISM is thought to be $P/k = 10^{3-4} {\rm cm}^{-3} $K (\cite{oey2004}), star forming regions just like the Cygnus region in general have higher pressure of the order of $P/k = 10^{5-6} {\rm cm}^{-3}$K (\cite{malhotra2001}). 
Thus the pressure inside the CSB is similar to that of outside, meaning that the CSB is at the end of its expanding phase.
Using these parameters, the total thermal energy content becomes 
 \begin{equation}
E_T = 3/2 \times P \times f \times V = 9 \times 10^{51}  {\rm ergs}
\label{energy}
\end{equation}
\citet{lozinskaya2001} performed detailed analysis of Cyg OB2 and estimated the luminosity of stellar winds as $L_{x} \sim (1-2) \times 10^{39}$ergs s$^{-1}$.
The wind mechanical luminosity over a Cyg OB2 lifetime of $\simeq 2-3 \times 10^6$ years is more than enough to produce and power-up the CSB.

Another possibility is that the CSB is a hypernova remnant. 
\citet{tsunemi2004} estimated the thermal  energy content of over a dozen SNRs. 
They showed that the thermal energy content of SNRs observed are $\sim 10^{49}$ erg
and it depends on $D_X^{-0.2}$ where $D_X$ is a diameter of SNR in X-ray. 
This suggests that when we compare typical young SNR ($D_X = 3$ pc) and old SNR ($D_X = 30$ pc), the difference of thermal energy content is about factor of $\sim 1.5$. When we adopt this to CSB ($D_X = 660$), the difference of thermal energy content is only about factor of $ \sim 3$.
In conventional SNe, the initial explosion energy is around $\sim 10^{51}$ erg, this indicates that about 1\% is
converted to thermal energy.
When we apply this to equation (\ref{energy}), the initial explosion energy becomes around $\sim 10^{54}$ erg,
which is very similar to the explosion energy of hypernova predicted by \citet{paczy1998}.
Additional fact to back up hypernova hypothesis is that the Cyg OB2 is not in a center of the CSB.
\citet{kiminki2007} measured the radial velocity of the Cyg OB2 and showed that the mean radial velocity of the Cyg OB2 
is about 10.3 km s$^{-1}$ therefore the Cygnus OB2 was off center $2-3 \times 10^6$ years ago.
It is unlikely that the stellar winds from off center source can create a circular bubble such as the CSB.
\citet{comeron2007} found a very massive runaway star from Cyg OB2.
They estimated that the star has $(70 \pm 15)\, {\rm M}\solar$ and an age of $\sim 1.6$ Myr.
If the progenitor star was a runaway star from the Cyg OB2,  and exploded near the center of the CSB,
this can solve a problem that the Cyg OB2 is not in the center of the CSB.

Regardless of its power source, the energy budget and the $N_{\rm H}$ of each region suggest that the CSB is most likely a single unity.

\section{Conclusion}
We used MAXI/SSC observation data and ROSAT PSPC data to obtain the X-ray spectrum of the CSB.
Good energy resolution of MAXI/SSC allowed us to obtain Fe, Ne, Mg emission lines for the first time, confirming that it is thin-thermal spectrum, it also gives us relative abundance of the CSB.
On the other hand, we check the ROSAT data to determine the precise value of $N_{\rm H}$.
The spectrum fit parameters of five different regions of the CSB show similar value of $N_{\rm H}$ and temperature indicating 
that the CSB is a single unity. 
Although calculated energy budget shows that the $2-3 \times 10^6$ years of stellar winds from Cygnus OB2 is enough to power up the CSB,
it is unlikely due to its off center position.
We conclude that the origin of the CSB is a hypernova rather than a combination of stellar winds from Cygnus OB2.

\section{Acknowledgments}
The authors thank to Dr K. Koyama for the careful reading of this manuscript.
This work is partly supported by JSPS KAKENHI Grant Number $(23000004)$.

\end{document}